\documentclass[prl,twocolumn,superscriptaddress,showpacs,floatfix,longbibliography]{revtex4-2}
\usepackage{mathrsfs,braket}
\usepackage{amssymb, amsbsy, amsmath, latexsym, dsfont, array, layout, graphicx,mathrsfs,color,ulem,bm,amsfonts}
\usepackage[colorlinks=true,citecolor=blue,urlcolor=blue,linkcolor=blue]{hyperref}

\begin{document}
\title{Chiral Dynamics of Ultracold Atoms under a Tunable SU(2) Synthetic Gauge Field}

\author{Qian Liang}
\thanks{These authors contributed equally to this work}
\author{Zhaoli Dong}
\thanks{These authors contributed equally to this work}
\affiliation{%
Zhejiang Province Key Laboratory of Quantum Technology and Device, School of Physics, and State Key Laboratory for Extreme Photonics and Instrumentation, Zhejiang University, Hangzhou 310027, China
}%
\affiliation{%
College of Optical Science and Engineering, Zhejiang University, Hangzhou 310027, China
}%
\author{Hongru Wang}
\affiliation{%
Zhejiang Province Key Laboratory of Quantum Technology and Device, School of Physics, and State Key Laboratory for Extreme Photonics and Instrumentation, Zhejiang University, Hangzhou 310027, China
}%
\affiliation{%
College of Optical Science and Engineering, Zhejiang University, Hangzhou 310027, China
}%
\author{Hang Li}
\author{Zhaoju Yang}
\affiliation{%
Zhejiang Province Key Laboratory of Quantum Technology and Device, School of Physics, and State Key Laboratory for Extreme Photonics and Instrumentation, Zhejiang University, Hangzhou 310027, China
}%
\author{Jian-Song Pan}
\affiliation{College of Physics and Key Laboratory of High Energy Density Physics and Technology of Ministry of Education, Sichuan University, Chengdu 610065, China}
\author{Wei Yi}
\email{wyiz@ustc.edu.cn}
\affiliation{CAS Key Laboratory of Quantum Information, University of Science and Technology of China, Hefei 230026, China}
\affiliation{CAS Center For Excellence in Quantum Information and Quantum Physics, Hefei 230026, China}
\author{Bo Yan}
\email{yanbohang@zju.edu.cn}
\affiliation{%
Zhejiang Province Key Laboratory of Quantum Technology and Device, School of Physics, and State Key Laboratory for Extreme Photonics and Instrumentation, Zhejiang University, Hangzhou 310027, China
}%
\affiliation{%
College of Optical Science and Engineering, Zhejiang University, Hangzhou 310027, China
}%

\date{\today}

\begin{abstract}
Surface currents emerge in superconductors exposed to magnetic fields, and are a key signature of the Meissner effect. Analogously, chiral dynamics were observed in quantum simulators under synthetic Abelian gauge fields. The flexible control of these simulators also facilitates the engineering of non-Abelian gauge fields, but their impact on the chiral dynamics remains elusive. Here, by employing the cutting-edge momentum-lattice technique, we implement a synthetic SU(2) gauge field in a spinful 1D ladder and study the rich chiral dynamics therein. We confirm the non-Abelian nature of the synthetic potential by observing the non-Abelian Aharonov-Bohm effect on a single plaquette. More importantly, the chiral current along the two legs of the ladder is observed to be spin-dependent and highly tunable through the parameters of the gauge potential. We experimentally map out different dynamic regimes of the chiral current, and reveal the underlying competition between overlaying flux ladders with distinct spin compositions. Our experiment demonstrates the dramatic impact of non-Abelian gauge fields on the system dynamics, paving the way for future studies of exotic synthetic gauge fields on the versatile platform of momentum lattices.
\end{abstract}

\maketitle
Gauge theory underlies our modern understanding of the quantum world: from interactions between elementary particles~\cite{Higgs1964prl,Kibble1967pr} to emergent gauge symmetries in strongly correlated matter~\cite{Senthil2000prb,Stromer1999rmp,Lee2008rpp}, gauge fields play a central role in a wealth of physical contexts. In recent years, the ever-growing ability to engineer synthetic gauge fields in quantum simulators offers fresh opportunities. On one hand, it is possible to emulate gauge fields that already exist in nature, now in the highly controllable synthetic settings such as ultracold atoms~\cite{lin2009nature,Aidelsburger2013prl,Miyake2013prl,fang2017sa,ABcage,Atala2014np}, superconducting qubits~\cite{roushan2017np}, as well as photons~\cite{Fang2012prl,fang2012np,Umucalilar2022pra,Mittal2014prl,schine2016nature} and phonons~\cite{Yang2017prl,mathew2020nn,chen2021prl,xiao2015np}. For instance, the experimental synthesis of the simplest Abelian gauge field, the scalar and vector potentials of electromagnetism, has led to the observation of chiral edge current of charge-neutral atoms in a flux ladder~\cite{Atala2014np}, a phenomenon analogous to the surface current in superconductors under the Meissner effect~\cite{Bardeen1955pr,Bardeen1957pr}. On the other hand, a rich variety of exotic gauge potentials become experimentally accessible~\cite{Dalibard2011rev,goldman2014review,Zhai2015rpp,Lin2011nature,wang2012prl,Huang2016np,wu2016science}. In particular, the realization of synthetic non-Abelian gauge fields underlies the recent surge of interest in simulating topological phenomena such as the non-Abelian geometric phase and topology~\cite{Liu2014prl,di2020nc,Wang2021science,Sun2018prl}, the (anomalous) quantum Hall effect~\cite{Wunderlich2005prl,Liang2023prr}, and the Yang's monopole~\cite{sugawa2018science,sugawa2021npj}. Therein, the non-Abelian gauge fields couple the internal and external degrees of freedom of the system, leading to intriguing geometric structures in the Hilbert space. Through these couplings, the non-Abelian gauge fields are also bound to affect the dynamic or transport properties of the system, but their impact remains largely unexplored.

\begin{figure}[tbp]
	\centering
	\includegraphics[width=0.48\textwidth]{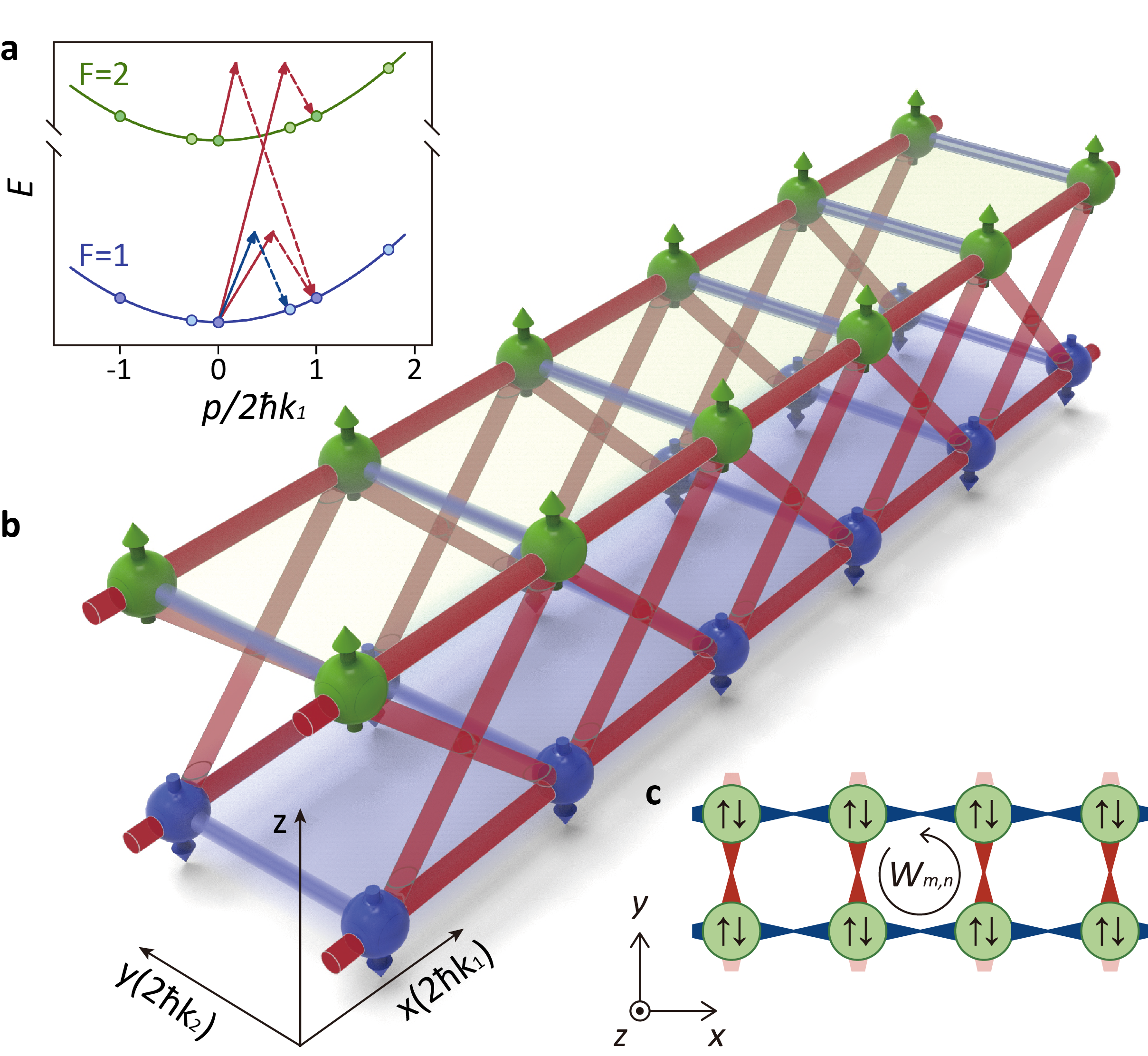}
	\caption{Synthesis of the SU(2) gauge fields in the RML. (a) Energy dispersions of the relevant ground-state hyperfine manifolds of $^{87}$Rb atoms. Some typical two-photon couplings are shown here: the red arrows indicate the two-photon transitions with the recoil momentum $2\hbar k_1$, and the blue arrows indicate those with $2\hbar k_2$. (b) Schematic illustration of the momentum lattice formed by the couplings in (a). The two internal hyperfine states are encoded as the spin-down ($F=1$) and spin-up ($F=2$) states. The color codes of the bonds and sites and consistent with those of the coupling processes and hyperfine states in (a). (c) A spin-1/2 ladder with an arbitrary SU(2) gauge field. Each site has spin-up and spin-down internal degrees of freedom. Here $n$ and $m$ respectively represent the index in the synthetic $x$ and $y$ direction.}\label{fig1}
\end{figure}

In this work, we experimentally study the effects of a highly tunable SU(2) gauge field on the chiral dynamics along a spin-$1/2$ 1D ladder. As illustrated in Fig.~\ref{fig1}, we encode the spin and lattice-site degrees of freedom into the combined synthetic dimensions of hyperfine and momentum states of a Bose-Einstein condensate of $^{87}$Rb atoms. Adopting a Raman momentum-lattice (RML) construction, we enforce the ladder geometry and gauge couplings through a series of appropriately designed Raman and Bragg couplings. The form and strength of the synthetic gauge field are conveniently tuned by manipulating the amplitude and phase of the coupling lasers. We first demonstrate the non-Abelian nature of the gauge couplings through the non-Abelian Aharonov-Bohm (AB) effect~\cite{naABprd, Dalibard2011rev, soljacic2019science, chen2019nc, huo2014sr}. Specifically, focusing on a single plaquette in the ladder, we sequentially switch on couplings to drive the atoms hopping around the plaquette in either the clockwise or counterclockwise direction. When the atoms come back to the starting position, we find that their final states are path-dependent in general, which can be understood as the non-Abelian counterpart of the well-known AB effect. We then systematically characterize the chiral dynamics by probing the chiral current, whose amplitude and direction are sensitive to the parameters of the gauge field. By decomposing the synthetic lattice into two sets of overlaying flux ladders, we reveal that the observed rich chiral dynamics derives from the competition between the flux ladders, as each set hosts a unique spin-dependent chiral current. Our experiment constitutes a natural starting point for studying the interplay of non-Abelian gauge fields, lattice geometry, disorder, and many-body interactions, all of which are independently tunable in the RML.

\begin{figure*}[tbp]
	\centering
	\includegraphics[width=0.8\textwidth]{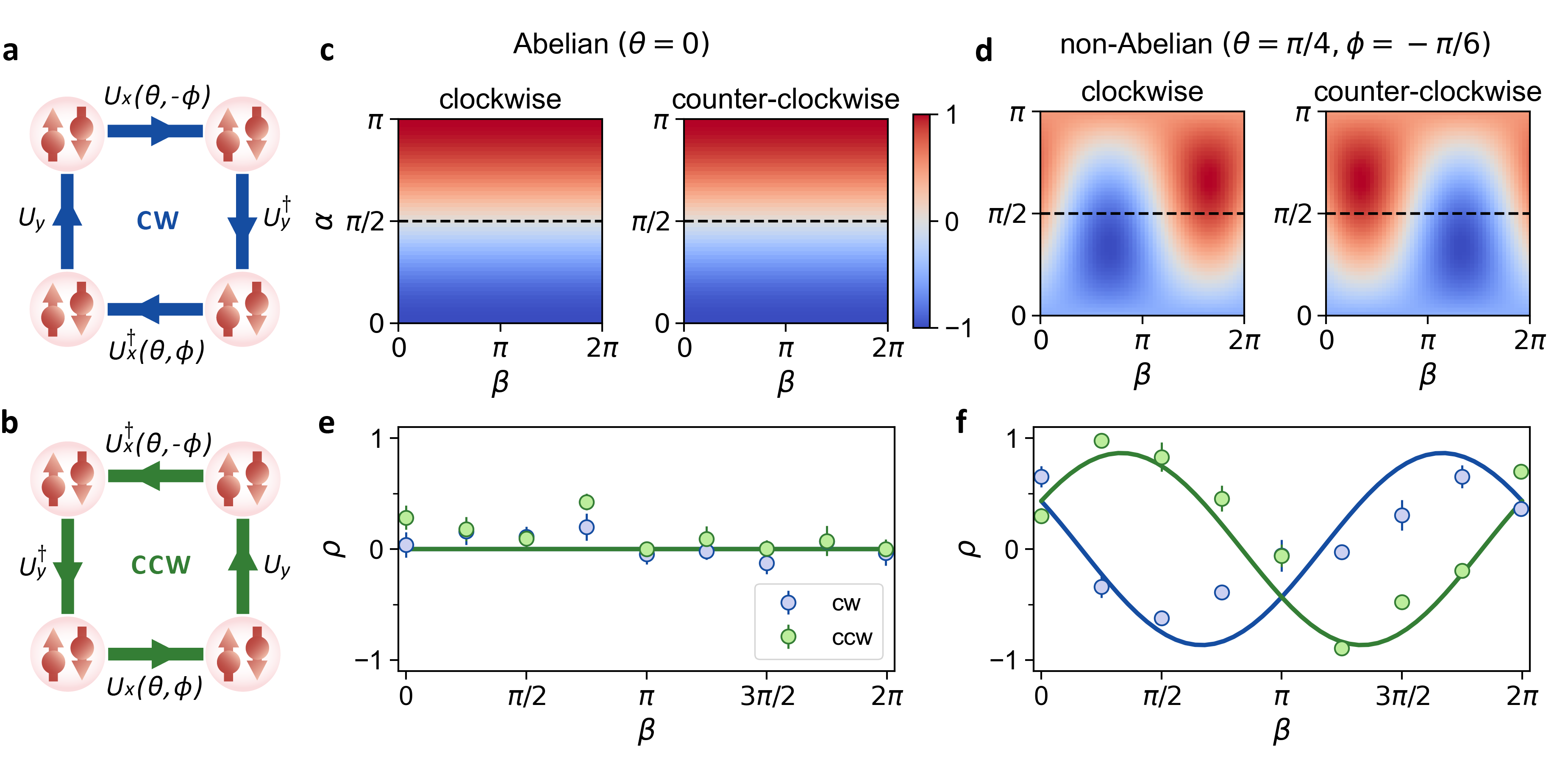}
	\caption{Schematic illustrations of the loop operators for the clockwise (a) and counterclockwise (b) rotations. Both operators start and end at the bottom-left corner of the plaquette. (c) and (d) show the numerically simulated results for the final states following the actions of $W_{\text{cw}}$ and $W_{\text{ccw}}$, respectively for the Abelian ($\theta=0,\varphi=0$) and non-Abelian cases ($\theta=\pi/4,\phi=-\pi/6, \varphi=0$). The experimental parameters are taken along the black dashed lines. (e) and (f) show the experimental results of the polarization ratio $\rho$ for different initial states characterized by various $\beta$ and a fixed $\alpha=\pi/2$. Blue (green) dots and lines are the corresponding experimental and numerical results for the clockwise (counter-clockwise) rotations. The error bars represent the standard deviation of the mean.
	}\label{fig2}
\end{figure*}

\begin{figure}[tbp]
	\centering
	\includegraphics[width=0.45\textwidth]{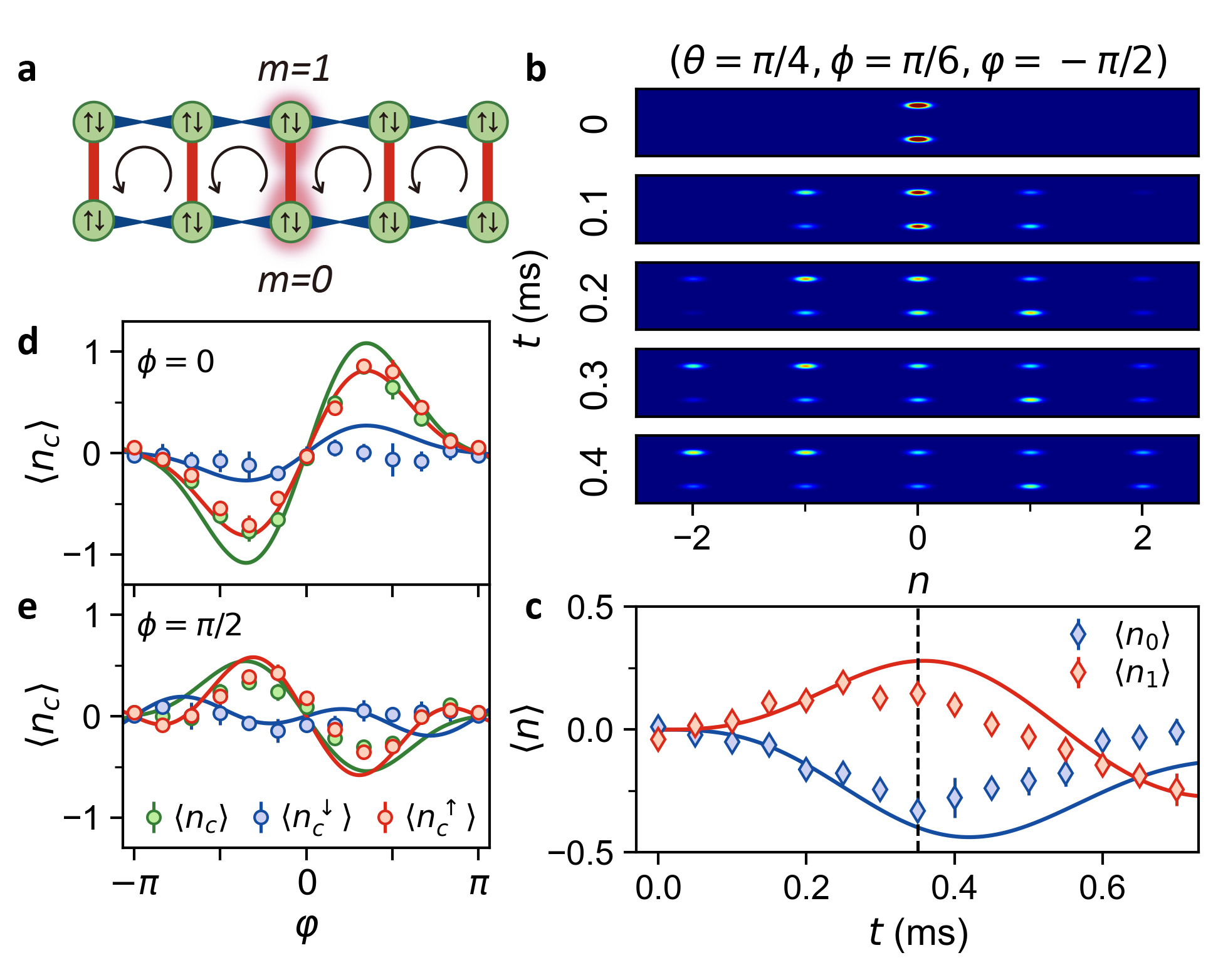}
	\caption{Spin-dependent chiral dynamics. (a) The spin-1/2 ladder in our experiment. The condensate is initially prepared in a superposition of sites $\{0,0\}$ and $\{0,1\}$ and in the spin-down states (shaded in red). (b) Time evolution of the condensate population with $\theta=\pi/4,\phi=\pi/6,\varphi=\pi/2$, and $J=h \times 0.50(2)~\text{kHz}$. (c) The average displacements $\langle n_{0,1}\rangle$ along the $m=0$ leg (red) and $m=1$ leg (blue), extracted from (b). The black dashed line indicates $t=0.35~\text{ms}$, as used in (d)(e). (d)(e) The chirality $\langle n_c\rangle$ versus $\varphi$ at $t=0.35~\text{ms}$ for $\theta=\pi/3, \phi=0$ (d) and $\theta=\pi/3, \phi=\pi/2$ (e), respectively. The blue (red) dots represent the chiral displacement of the spin-down (spin-up) states, while the green dots are the overall chiral displacement. In (c-e), diamonds (or dots) are the experiment results and the solid lines are from numerical simulations. Error bars stand for the standard deviation of the mean.
	}\label{fig3}
\end{figure}	

{\bf Experimental implementation and model Hamiltonian.}
We experimentally implement the RML following the schematics illustrated in Fig.~\ref{fig1}. Two sets of Bragg lasers, with wavelengths $\lambda_1=794.7~\text{nm}$ and $\lambda_2=1064~\text{nm}$ respectively, couple discrete momentum states in the $F=1$ and $F=2$ ground-state hyperfine manifolds. The coupled discrete momenta are $2n\hbar k_1$ along the synthetic $x$ axis, and $2m\hbar k_2$ along the $y$ axis, respectively, where $k_{1,2}=2\pi/\lambda_{1,2}$, and $n,~m\in \mathbb{Z}$. To couple states in different hyperfine manifolds, we introduce Raman-Bragg processes, as illustrated in the inset of Fig.~\ref{fig1}(a). Mapping different momentum states to lattice sites in the two-leg ladder (labeled by $\{n,m\}$), and the two internal hyperfine degrees of freedom to the spins, we thus realize a spin-$1/2$ ladder in the synthetic dimensions of the atoms. By adjusting the parameters of the coupling lasers, a tunable SU(2) lattice gauge field is achieved, as depicted in Fig.~\ref{fig1}(c)~\cite{supp}.

The system can be described by the following effective Hamiltonian
\begin{align}
H = \sum_{n,m}J(\hat{a}^{\dagger}_{n+1,m} U_m^{(x)} \hat{a}_{n,m} + \hat{a}^{\dagger}_{n,m+1} U_n^{(y)} \hat{a}_{n,m}) + \rm{H.c}, \label{eqH}
\end{align}
where $J$ is the laser-induced tunneling amplitude,  $\hat{a}_{n,m}=(\hat{a}_{n,m,\uparrow},\hat{a}_{n,m,\downarrow})^T$ and $\hat{a}_{n,m}^\dagger=(\hat{a}^\dagger_{n,m,\uparrow},\hat{a}^\dagger_{n,m,\downarrow})^T$ are the spinor annihilation and creation operators on the $\{n,m\}$ site, respectively. The hopping between two adjacent sites along the $x$ axis is accompanied by a spin-state rotation given by $U^{(x)}$, with
\begin{align}
	U_{x}(\theta,\phi) = \begin{pmatrix} \cos\theta & i\sin\theta e^{i\phi} \\ i\sin\theta e^{-i\phi} & \cos\theta \end{pmatrix}.
\end{align}
In our experiment, we take $U^{(x)}_0=U_x(\theta,\phi)$ and $U^{(x)}_1=U_x(\theta,-\phi)$ by setting the coupling parameters. Along the $y$ axis, we impose $U^{(y)}_n=U_y(n\varphi)=e^{in\varphi} \mathbb{I}$, where $\mathbb{I}$ is an identity $2\times 2$ matrix. While $U_x$ and $U_y$ conspire to give a synthetic SU(2) gauge field, both the form and strength of the gauge field can be easily controlled by adjusting the parameters of $\theta, \phi$, and $\varphi$. Under the condition $\{\theta=n\pi\}$, or $\{\phi=m\pi\}$, or $\{\theta=(2n+1)\pi/2\,\, \text{and}\,\,\phi=(2m+1)\pi/2\} ~(n,m \in \mathbb{N})$, the gauge field becomes Abelian. Hence, by controlling the coupling-laser parameters, the Abelian or non-Abelian nature of the gauge field can be tuned conveniently~\cite{supp}.

\begin{figure*}[tbp]
	\centering
	\includegraphics[width=0.9\textwidth]{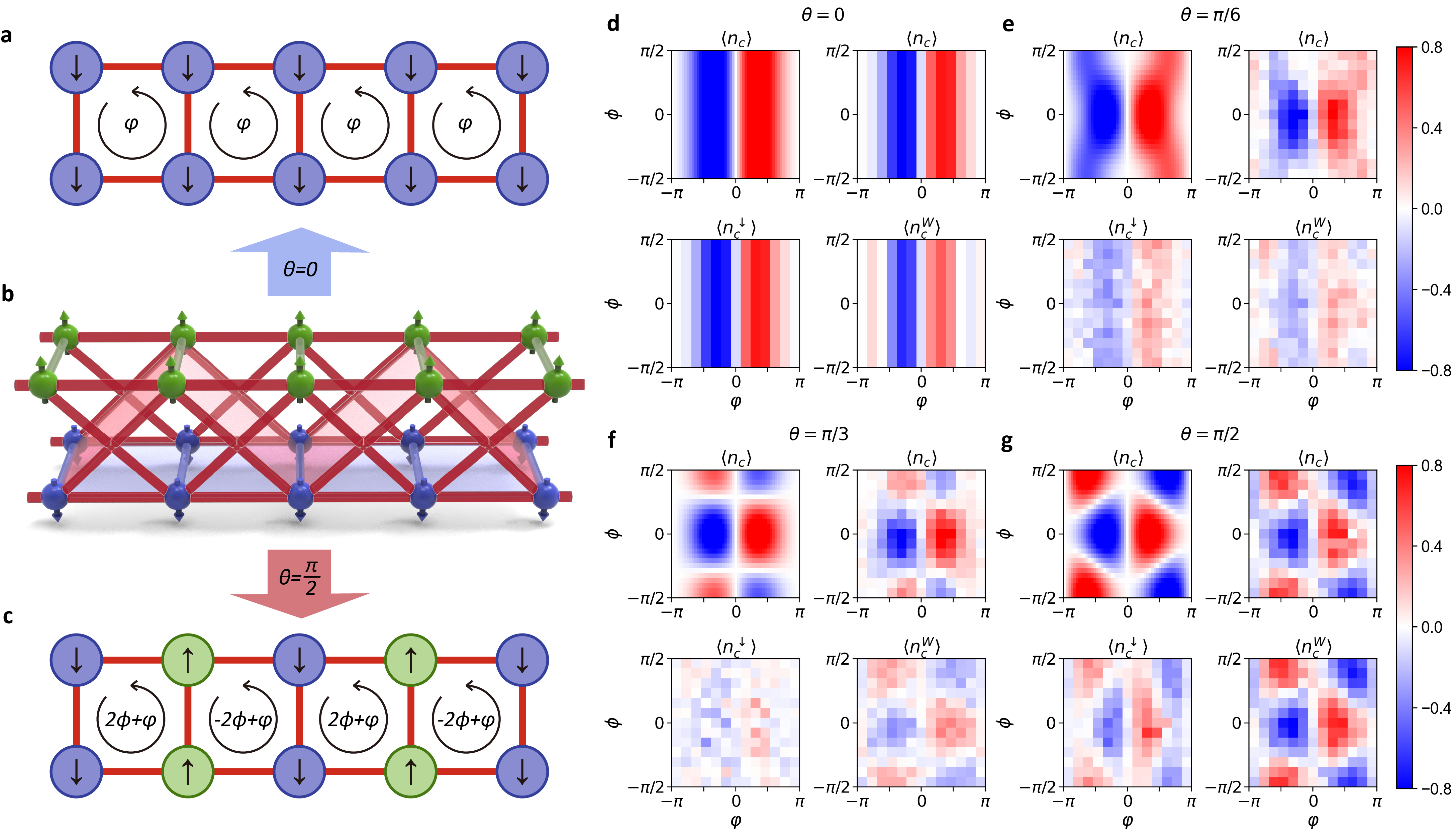}
	\caption{Dynamic regimes of the chiral current. (a-c) The SU(2) ladder in (b) can be decomposed into two sets of overlaying flux ladders. When $\theta=0$, the SU(2) ladder is reduced to a pair of decoupled flux ladders with different spin species. We illustrate the spin-down ladder in (a), corresponding to the blue-shaded horizontal plane in (b). When $\theta=\pi/2$, the SU(2) ladder is reduced to a decoupled pair of zig-zag ladders with alternating spins on adjacent sites. We illustrate one of them in (c), which is shaded in red in (b). (d-g) show the chiral dynamic regimes in the plane of $\phi$ and $\varphi$, under different values of $\theta$. For each panel, the upper-left subfigure displays the numerical result of $\langle n_c\rangle$, and the upper-right one shows the corresponding experimental result. The lower subfigures are experimental data for $\langle n^\downarrow_c \rangle$ and $\langle n^W_c\rangle$, respectively. The tunneling strength is set at $J=h\times 0.50(2)~\text{kHz}$, and the time of measurement is fixed at $t=0.35~\text{ms}$ for all cases.}\label{fig4}
\end{figure*}
	
{\bf Non-Abelian AB effect.}
Under the AB effect, a charged particle moving around a closed loop acquires a geometric phase as a result of the U(1) gauge potential along the path, which is responsible for the magnetic flux penetrating the loop. However, the situation is quite different when the particles are subjected to a non-Abelian gauge field. To see this, we focus on a single plaquette, and define the loop operators of the clockwise (cw) or counterclockwise (ccw) paths, following the effective Hamiltonian (\ref{eqH})
\begin{align}
W_{\text{cw}} &= U_x^\dagger(\theta,\phi)U_y^\dagger((n+1)\varphi) U_x(\theta,-\phi)U_y(n\varphi),\\
W_{\text{ccw}}&=U_y^\dagger(n\varphi) U_x^\dagger(\theta,-\phi) U_y((n+1)\varphi) U_x(\theta,\phi).
\end{align}
Under an Abelian gauge potential, $W_{\text{cw}}=e^{-i\varphi}\mathbb{I}$ and $W_{\text{ccw}}=e^{i\varphi}\mathbb{I}$. Atoms going around a closed loop via different paths would accumulate geometric phases with opposite signs. In the non-Abelian scenario, however, the two loop operators are different: atoms going around the plaquette undergo an internal-state rotation that is path-dependent.

For our experiment, we initialize the atoms in site $|n=0,m=0\rangle$ (referred to as $|0,0\rangle$ in the following) with variable spin states
\begin{align}
|\psi_{\text{ini}}\rangle = \left(i\sin\frac{\alpha}{2}e^{i\beta}\left|\uparrow\right\rangle+ \cos\frac{\alpha}{2}\left|\downarrow\right\rangle\right) \otimes |0,0\rangle.
\end{align}
Here $\alpha$ determines the spin composition and $\beta$ is the relative phase. We employ a sequence of four $\pi$-pulses to implement the loop operators. The final states are denoted as $|\psi^{\text{cw}}_{\text{fin}}\rangle=W_{\text{cw}}|\psi_{\text{ini}}\rangle$ and $|\psi^{\text{ccw}}_{\text{fin}}\rangle=W_{\text{ccw}}|\psi_{\text{ini}}\rangle$ for the clockwise and counter-clockwise loops, respectively. To characterize the internal-state response, we define the polarization ratio  $\rho=|\langle \uparrow|\psi_{\text{fin}} \rangle|^2 - |\langle\downarrow|\psi_{\text{fin}}\rangle|^2$. Under an Abelian gauge field, $\rho^\text{cw}=\rho^\text{ccw}$, regardless of the initial state [as shown in Fig.~\ref{fig2}(c)]. Whereas under a non-Abelian gauge potential,  $\rho^\text{cw}\neq\rho^\text{ccw}$ for most initial states [see Fig.~\ref{fig2}(d)]. These results are confirmed by the experimental observation for $\alpha=\pi/2$. In Fig.~\ref{fig2}(e), we set $\theta=0, \varphi=0$, and observe $\rho^\text{cw}=\rho^\text{ccw}=0$ for both the clockwise and counter-clockwise loops. Conversely, under a non-Abelian gauge field, the polarization ratio $\rho$ exhibits an apparent dependence on the path direction, as well as on the relative phase $\beta$ of the initial state [see Fig.~\ref{fig2}(f)].

{\bf Chiral dynamics under the Non-Abelian gauge field.}
We are now in a position to discuss the dynamic consequence of the SU(2) gauge field. In previous studies of flux ladders, an outstanding feature is the chiral current: particles flow toward opposite directions along the two legs, with the overall chirality dependent on the sign of the flux. Here we show that, when the form of the gauge potential is continuously tuned, both the amplitude and direction of the chiral current are significantly modified, as the system exhibits multiple dynamic regimes.

We experimentally study the chiral dynamics on a 1D ladder of $2\times5$ sites. As illustrated in Fig.~\ref{fig3}(a), this corresponds to $m\in\{0,1\}$ and $n\in\{-2,-1,0,1,2\}$. The condensate is initialized in the spin-down state, with $|\psi(t=0)\rangle=\left|\downarrow\right\rangle\otimes\frac{\sqrt{2}}{2}(|0,0\rangle + |0,1\rangle)$. Under a typical non-Abelian gauge field with $(\theta=\pi/4, \phi=\pi/6, \varphi=\pi/2)$, the atomic flow shows a shearing behavior, as shown in Fig.~\ref{fig3}(b). As time evolves, the atom population displays a directional propagation in both legs, where atoms in the $m=0$ leg flow to the left and those in the $m=1$ leg to the right.

To better visualize the directional flow, we define the average displacement for $m$-th leg
\begin{align}
\langle n_m(t)\rangle = \sum_n n(|\psi_{n,m,\uparrow}(t)|^2 + |\psi_{n,m,\downarrow}(t)|^2),
\end{align}
where $\psi_{n,m,\sigma}(t)=\langle\psi(t)|\sigma\rangle\otimes|n,m\rangle$ ($\sigma\in\{\uparrow,\downarrow\}$). The corresponding experiment results are shown in Fig.~\ref{fig3}(c), where atomic currents are opposite in direction on the two legs. Note that for $t\gtrsim 0.6$ ms, the atoms reach the lattice boundary, and the directional flow diminishes. Based on this observation, we characterize the chirality of the flow using the chiral displacement $\langle n^{\sigma}_c\rangle=\langle n^{\sigma}_1\rangle -\langle n^{\sigma}_0\rangle$ at $t=0.35$ ms, when the atoms are still in the bulk. The presence of chirality in spin component $\sigma$ is reflected in a non-vanishing $n^{\sigma}_c$, with the chirality given by its sign. Likewise, the overall chirality in the atomic density flow is given by $\langle n_c\rangle=\langle n^{\uparrow}_c\rangle+\langle n_c^{\downarrow}\rangle$.

In Fig.~\ref{fig3}(d), we show the measured $n_c$ and $n_c^\sigma$ for $\theta=\pi/3,\phi=0$, where the gauge field is reduced to an Abelian form. The synthetic flux through each plaquette is given by $\varphi$. Consistent with previous results of a flux ladder~\cite{Atala2014np,fang2017sa}, we observe $\langle n_c\rangle >0$ ($\langle n_c\rangle <0$) for $\varphi>0$ ($\varphi<0$), indicating the change of chirality as the flux is reversed. However, when $\theta=\pi/3,\phi=\pi/2$, as shown in Fig.~\ref{fig3}(e), $\langle n_c\rangle$ is reversed compared to that in Fig.~\ref{fig3}(d), and the chiral current is dominated by that of the spin-down species. This suggests that, under the non-Abelian gauge field, chiral dynamics becomes spin-dependent, and the overall chirality can be flipped.

To understand the rich chiral dynamics of the system, let us examine two limiting cases. First, when $\theta=0$, the gauge field becomes Abelian, and the two spin components are decoupled, as the system is reduced to two independent flux ladders [see Fig.~\ref{fig4}(a)(b)]. In this case, the measured chiral dynamics depends only on $\varphi$, consistent with that of a simple flux ladder [see Fig.~\ref{fig4}(d)]. Second, when $\theta=\pi/2$, the system is again reduced to two decoupled flux ladders, each consisting of alternating spin species along the ladder [see Fig.~\ref{fig4}(b)(c)]. For a condensate initially prepared in the spin-down state, the ensuing dynamics only involves a zig-zag flux ladder composed of the red-shaded plaquettes in Fig.~\ref{fig4}(b). To characterize the chirality of the dynamics on such a zig-zag ladder, we define
\begin{align}
	\langle n_c^\text{w}\rangle &= \sum_{n} [2n|\psi_{2n,1,\downarrow}|^2+(2n+1)|\psi_{2n+1,1,\uparrow}|^2 \nonumber \\
	&-2n|\psi_{2n,0,\downarrow}|^2-(2n+1)|\psi_{2n+1,0,\uparrow}|^2],
\end{align}
As shown in Fig.~\ref{fig4}(g), multiple distinct dynamic regimes emerge in the parameter space of $\varphi$ and $\phi$. There are two kinds of boundaries here. One is given by $(\varphi=0,\pi)$, where the flux in Fig.~\ref{fig4}(c) becomes staggered. In this case, the chiralities of adjacent plaquettes cancel out~\cite{Dhar2013prb,Sachdeva2018pra}, leading to the corresponding boundaries observed in Fig.~\ref{fig4}(g). The rest of the boundaries are pinned by the points $(\phi=0,\varphi=\pm\pi)$ and  $(\phi=\pm\pi/2,\varphi=0)$, where the chirality flips as the fluxes cross $\pi$ or $-\pi$. 

Finally, for $\theta$ taking values in between $0$ and $\pi/2$, the chiral dynamics can be understood as the result of the competition between the two sets of flux ladders. Indeed, Fig.~\ref{fig4}(e)(f) show the transition from one limit to the other. The boundary of $\varphi=n\pi$ persists for all values of $\theta$, while $\phi$ plays an increasingly important role as $\theta$ changes from $0$ to $\pi/2$, giving rise to the rich chiral dynamics under a general SU(2) gauge field.
	
{\bf Discussion.}
To summarize, we demonstrate the spin-dependent chiral dynamics in a 1D ladder under an SU(2) synthetic gauge field. The flexibility of our Raman momentum-lattice construction provides opportunities to study synthetic non-Abelian gauge fields in higher dimensions, or topological models and the associated topological phenomena where various forms of the non-Abelian gauge fields play a key role. These include the non-Abelian Hofstadter model~\cite{Dalibard2011rev, Osterloh2005prl, Goldman2009pra, soljacic2019science, yang2020light}, the higher-order topological insulators~\cite{goldman2014review,Zamora2011pra,di2020nc}, and the quantum (spin) Hall effect~\cite{Goldman2010prl}. It also builds a foundation for studying the interplay of many-body interaction and non-Abelian gauge fields, with the exciting possibility of probing intrinsic topological order and fractional excitations on the highly tunable platform of momentum lattices.

\begin{acknowledgments}
{\it Acknowledgement:}
We acknowledge the support from the National Key Research and Development Program of China under Grant No.2022YFA1404203 and No. 2023YFA1406703, The National Natural Science Foundation of China under Grant No. U21A20437, No. 12074337, No. 11974331, and No. 12374479, Natural Science Foundation of Zhejiang Province under Grant No. LR21A040002, Zhejiang Province Plan for Science and Technology Grant No. 2020C01019, and the Fundamental Research Funds for the Central Universities under Grant No. 2021FZZX001-02 and 226-2023-00131, the China Postdoctoral Science Foundation under Grant No. 2023M733122.
\end{acknowledgments}
	
\bibliographystyle{apsrev4-2}
\bibliography{non-abelian}
\end{document}